\documentclass[showpacs,twocolumn,pre,floatfix]{revtex4}
\usepackage{graphicx}
\usepackage{times}
\usepackage{bm}
\usepackage{amsmath}
\DeclareMathOperator{\tr}{tr}
\DeclareMathOperator{\diag}{diag}
\DeclareMathOperator{\re}{Re}
\DeclareMathOperator{\im}{Im}
\DeclareMathOperator{\ai}{Ai}
\DeclareMathOperator{\bi}{Bi}

\begin{document}

\date{April 2002}
\title{Statistics of finite-time Lyapunov exponents
in a random time-dependent potential}

\author{H. Schomerus}
\email{henning@mpipks-dresden.mpg.de}
\author{M. Titov}
\email{titov@mpipks-dresden.mpg.de}
\affiliation{Max-Planck-Institut f\"ur Physik komplexer Systeme,
N\"othnitzer Str. 38, 01187 Dresden, Germany
}

\begin{abstract}
The sensitivity 
of trajectories over finite time intervals $t$
to perturbations of the initial conditions
can be associated with a finite-time
Lyapunov exponent $\lambda$, obtained from the elements $M_{ij}$
of the stability matrix $M$. For globally chaotic dynamics
$\lambda$ tends to a unique value (the usual
Lyapunov exponent $\lambda_\infty$) as  $t$ is sent to infinity, but 
for finite $t$ it depends 
on the initial
conditions of the trajectory and can be considered as a statistical
quantity.
We compute for a particle moving in a random time-dependent potential
how the distribution function $P(\lambda;t)$
approaches 
the limiting
distribution $P(\lambda;\infty)=\delta(\lambda-\lambda_\infty)$.
Our method also applies to the tail of the distribution,
which determines
the growth rates of positive  moments of $M_{ij}$.
The results are also applicable to the problem of wave-function
localization in a disordered one-dimensional potential.
\end{abstract}

\maketitle

\section{Introduction}

In this work,
we give a uniform description of the complete asymptotic statistics of
the finite-time Lyapunov exponent  for a particle moving in a random time-dependent potential.
The Lyapunov exponent $\lambda_{\infty}$
characterizes the sensitivity of trajectories 
to small perturbations
of the initial conditions and plays a fundamental role
in the characterization of systems which display deterministic chaos \cite{ott}.
The Lyapunov exponent
is defined in the joint limits of vanishing initial perturbation
and infinitely large times. 
In a hyperbolic Hamiltonian system  $\lambda_\infty$
may be obtained from any non-periodic trajectory,
because for arbitrarily long times 
the trajectories uniformly explore the complete phase space.

A widely studied generalization of $\lambda_\infty$ is the
finite-time Lyapunov exponent
\cite{ott,fujisaka,grassberger0,Badii,tel,grassberger,sepulveda,Vaienti,%
romeiras,vulpiani,amitrano,%
eckhardt,ernst,adrover,Prasad%
,Diakonos,yamada},
which is defined for finite stretches
(time interval $t$)
of trajectories
(generalizations to finite perturbations also exist \cite{aurell}).
The sensitivity  of the dynamics
to initial perturbations
is given by
the stability matrix map $M$, which is the linearization of the
map of initial coordinates to final coordinates.
In terms of elements $M_{ij}$ of $M$, the
(complex) finite-time Lyapunov exponent
may then be defined as
\begin{equation}
\lambda=\frac{1}{t}\ln M_{ij}.
\label{eq:ldef}
\end{equation}

In contrast to $\lambda_\infty$, $\lambda$
is not a unique number independent of the initial conditions,
but a fluctuating quantity  
with a distribution function $P(\lambda;t)$ (defined by uniformly
sampling all initial conditions in phase space).
This distribution function
determines, e.\,g.,
the generalized entropy and dimension spectra of dynamical systems
\cite{ott},
and more practically
the weak-localization correction to the conductance \cite{al1}
and the shot-noise suppression \cite{al2,oberholzer} in mesoscopic systems.
Finite-time Lyapunov exponents
also determine the wavefront stability of acoustic
and electromagnetic wave propagation through a random medium,
in the ray-acoustics/ray-optics regime
of short wave lengths (for a
recent application see Refs.\ \cite{wolfson,brown}).
Moreover, they have shifted
into the focus of attention due to 
recent advances in the understanding of the role of the Lyapunov
exponents for quantum-chaotic wave
propagation \cite{jalabert,cucchietti01a,cerruti,prosen,jacquod,cucchietti01,jacquod02}:
It has been observed that under certain conditions the Lyapunov exponent
can be extracted from the
decay of the overlap of two
wavefunctions which are propagated by two slightly different Hamiltonians (the so-called Loschmidt
echo). Since the overlap is studied as a function of time, the
distribution of the finite-time Lyapunov exponent is
directly relevant for these investigations. This extends also to
related semiclassical time scales, 
like to the Ehrenfest time 
$\sim (\log\hbar)/\lambda$, which is a semiclassical estimate of the 
diffraction time of wave packets due to the
chaotic classical dynamics. 

In the limit of infinite time $t$ the distribution function $P(\lambda;t)$
in a completely chaotic phase space
tends to the limiting form $P(\lambda;\infty)=\delta(\lambda-\lambda_\infty)$.
For large but
finite $t$ the bulk of the distribution function can be approximated
by a Gaussian centered around $\lambda_\infty$,
with the width vanishing $\propto t^{-1/2}$ as $t\to\infty$.
However,
many of the properties
determined by $P(\lambda;t)$
(like the generalized entropy and dimension spectra)
cannot be
calculated from the Gaussian bulk of the distribution function
\cite{ott}.

In this paper we investigate for
a particle moving in a one-dimensional random time-dependent potential how
$P(\lambda;t)$ approaches the limiting
distribution function
$P(\lambda;\infty)=\delta(\lambda-\lambda_\infty)$ for large times.
Our approach uniformly applies both to the bulk as well as to the 
far tail $\lambda \gg \lambda_\infty$ of the distribution function.
We find that the cumulant-generating function of $P(\lambda;t)$,
\begin{equation}
\eta(\xi )=\ln\left\langle \exp(\xi t \lambda) \right\rangle
=\sum_{n=1}^\infty\langle\langle \lambda^n\rangle\rangle\frac{(\xi t)^n}{n!}
,
\label{eq:cumuldef}
\end{equation}
(where the average $\langle \cdot \rangle$ is over initial
conditions and $\langle \langle \cdot \rangle \rangle$
denotes the cumulants),
takes the asymptotic form
\begin{equation}
\eta(\xi)=  \mu(\xi) t / t_c +{\cal O}(t^0)
,
\label{eq:res}
\end{equation}
with $\mu(\xi)$ a universal function (within the statistical model) and
\begin{equation}
t_c= \lambda_\infty^{-1} \mu^{(1)}
\label{eq:tc}
\end{equation}
a system-specific time-scale which can be determined from the
infinite-time-Lyapunov exponent and the constant
$ \mu^{(1)}=d\mu/d\xi|_{\xi=0}$
(by definition, $d\eta/d\xi|_{\xi=0}=\lambda_\infty t$).
The function $\mu(\xi)$ is given by the leading eigenvalue of a 
second-order differential equation in which $\xi$ appears as a parameter.
This eigenvalue can be calculated
perturbatively in $\xi$, which generates the
cumulants of $\lambda$. The values of $\mu$ at integer $\xi$
determine the asymptotic growth rates
$(1/t)\ln \langle M_{ij}^\xi\rangle= \mu(\xi) / t_c $
of positive moments of elements of the stability matrix.
We find that these values are given by
the leading eigenvalue of finite-dimensional matrices.

A random time-dependent potential
is often considered as a statistical model for 
the ergodic properties of hyperbolic chaotic motion, in the spirit
of the early work of Chirikov \cite{chirikov}.
The time dependence of the potential may be considered to mimic
the dependence of the potential in the eigentime along the trajectory.
In the context of finite-time Lyapunov
exponents there have been indications that a
statistical description is usually valid for the chaotic background of its distribution
\cite{Diakonos},
while system-specific deviations may exists in some exceptional cases even in the bulk of the
distribution function \cite{Prasad}.
While the statistical model considered in this work
is tailored to a specific class of Hamiltonian systems,
it can be modified
straightforwardly to other classes of chaotic systems (this is briefly
described at the end of
this paper).

The problem of finite-time Lyapunov exponents in the random
time-dependent potential is equivalent to the problem of
wave-localization in a random one-dimensional potential
\cite{Anderson,Borland,Landauer,Mott,reviews}, because the
equations of motion for the matrix elements $M_{ij}$ are formally 
equivalent to the corresponding Schr{\"o}dinger equation
\cite{vulpiani,crisanti}. 
Indeed, the  Fokker-Planck equation employed in this work is based on the phase
formalism 
described, e.g., in Ref.\
\cite{Frisch,Halperin,lifshitz}.
Hence, the asymptotic statistics of
the finite-time Lyapunov exponent presented in this work
directly is of interest and can
be transferred to this field of research.
A number of additional areas of
application of our method come into scope if one considers
the vast arena
of problems which can be analyzed by products of random matrices,
since the finite-time Lyapunov exponents
are a valuable way to characterize the eigenvalues of these products
\cite{vulpiani}.

The plan of this paper is as follows:
In Sec.\ \ref{sec:form} we formulate the problem of finite-time Lyapunov
exponents in the one-dimensional random time-dependent potential.
In Sec.\ \ref{sec:sol} we show how the cumulant-generating function can
be related to the parameterized eigenvalue of a second-order
differential equation, and that the cumulants can be calculated systematically.
Positive moments of $M_{ij}$ are calculated in   Sec.\ \ref{sec:pos}.
We close the paper with
discussion and conclusions in  Sec.\ \ref{sec:dis}.

\section{\label{sec:form}Formulation of the problem}

\subsection{Statistical model}

Let us consider a time-dependent Hamiltonian system with one degree of
freedom (canonically conjugated 
coordinates $x$, $p$), and the Hamiltonian given by
\begin{equation}
H = \frac{p^2}{2m}  +V(x,t) + \frac{V_2}{2} x^2.
\label{eq:ham}
\end{equation}
Here $V(x,t)$ is a time-dependent potential and $m$ is a mass.
We also allow for an additional static potential with curvature $V_2$ acting in the background
of the random potential (this potential is repulsive for $V_2<0$ and attractive for $V_2>0$).

We introduce the map ${\cal F}_t$
which propagates initial conditions 
$(x_i,p_i)$ over a time interval $t$ to the final coordinates
$(x_f,p_f)={\cal F}_t(x_i,p_i)$.
The  stability matrix $M$ is the linearization of the map ${\cal F}_t$
and describes the sensitivity of the final coordinates to 
a small perturbation of the initial conditions,
\begin{equation}
M=\frac{\partial (x_f,p_f)}{\partial (x_i,p_i)} = \left(\begin{array}{cc} M_{11}
& M_{12} \\ M_{21} & M_{22}
\end{array} \right).
\end{equation}
Area preservation of the dynamics in phase space entails the property
$\det M=1$ of the stability matrix.

We are interested in the evolution of the stability matrix with
given initial conditions and
increasing time interval $t$.
According to Hamilton's equations of motion
the
stability matrix
fulfills the differential equation
\begin{equation}
\frac { dM}{dt}=K M,
\quad
K=\left(\begin{array}{cc} 0 & m^{-1} \\ v & 0 \end{array}
\right), 
\label{eq:dm}
\end{equation}
where the function $v(t)$ in the matrix $K$ is given by
\begin{equation}
v = -V_2 - \left. \frac{d^2 V}{d x^2}\right|_{(x,p)=(x_f,p_f)}.
\end{equation}
This differential equation is supplemented by the initial conditions 
\begin{equation}
\quad M(0)=\diag(1,1),
\label{eq:initial}
\end{equation}
corresponding to the identification of 
the initial and final coordinate systems for $t=0$.

In order to study the statistical behavior of the stability matrix
we now assume that
$v(t)$ is
a randomly fluctuation function equivalent
to Gaussian random $\delta$-correlated
noise, 
\begin{equation}
\langle v(t) \rangle =-V_2,\qquad
\langle v(t_1) v(t_2) \rangle =2D\delta(t_1-t_2).
\label{eq:v}
\end{equation}
The condition of a vanishing mean of the time-dependent part of $v$
corresponds to the observation that
the incidence of
positive and negative curvature of the potential
landscape along a typical chaotic trajectory should be identical.
The $\delta$-function correlations are
valid if the correlation time of the fluctuations
is smaller than the mean free
transport time in the random potential. 
The constant $D$ (similar to a diffusion constant, but not identical with
conventional
diffusion constants of motion in phase space) can be related to 
the strength of the temporal fluctuations
of the potential $V(x,t)$.  
However, both $D$ as well as the mass $m$
can be eliminated from the subsequent
analysis by rescaling quantities in the following way:
\begin{eqnarray}
&&t=t_c t',\qquad v=(D/m)t_c v',\qquad V_2=(D/m)t_c V_2'\nonumber\\
&& M_{12}=(t_c/m)M_{12}',\qquad M_{21}=(m/t_c)M_{21}', \nonumber \\
&&  M_{11}=M_{11}',\qquad M_{22}=M_{22}'.
\label{eq:resc}
\end{eqnarray}
Here we defined the characteristic time scale
\begin{equation}
t_c=m^{2/3}D^{-1/3}.
\label{eq:tc2}
\end{equation} 
[In the course of our analysis we will see that this time scale also can be found from
Eq.\ (\ref{eq:tc}).]
The rescaled (primed) quantities fulfill Eqs.\ (\ref{eq:dm}), 
(\ref{eq:initial}), (\ref{eq:v})
with $D=m=1$.
Also note that the rescaling
leaves the property $\det M=1$ invariant.

\subsection{\label{sec:loc}Relation to one-dimensional localization}
The set of linear first-order differential equations (\ref{eq:dm}) can be decoupled by 
converting them into second-order differential equations. It is useful
to note (as mentioned in the introduction)
that the equations for
the elements $ M_{11}$ and $M_{12}$
are equivalent to the Schr{\"o}dinger equation, at energy $E=V_2/m$,
of a particle of mass $\hbar^2/2$ in a one-dimensional random
potential $(v+V_2)/m$ (of vanishing mean), with $t$ playing the role of the spatial coordinate,
\begin{equation}
\frac{d^2 M_{11}}{dt^2}=\frac{v}{m} M_{11},\quad
\frac{d^2 M_{12}}{dt^2}=\frac{v}{m} M_{12},
\label{eq:m11eq}
\end{equation}
while the other matrix elements are directly related to them by
\begin{equation}
M_{21}=m\frac{d M_{11}}{dt},\quad M_{22}=m\frac{d M_{12}}{dt}.
\label{eq:m2112}
\end{equation}
The problem of finite-time Lyapunov exponents hence is closely related to
the problem of one-dimensional localization
in a random potential, in which the Lyapunov exponent corresponds to the inverse decay
length of the wave function.

\section{\label{sec:sol}Cumulants of the finite-time Lyapunov exponent}

We now solve the problem of finding the 
probability distribution function 
of matrix elements $M_{ij}$ within 
the statistical model of chaotic dynamics, defined by the evolution
equation (\ref{eq:dm}) for $M$, with initial condition
(\ref{eq:initial}),  
and the statistical properties (\ref{eq:v}) of the random function $v$.
For the sake of definiteness
we will consider in this section the statistics of 
the upper diagonal element $M_{11}$.
The results directly carry over to
the other elements of $M$, as is discussed in
Sec.\ \ref{sec:equiv}.

\subsection{Cumulant-generating function as an eigenvalue}
We introduce the quantities 
\begin{equation}
u=\ln M_{11}',\quad z= \frac{M_{21}'}{M_{11}'},
\label{eq:u}
\end{equation}
where the relation
$u=\lambda t$ 
to the finite-time Lyapunov exponent $\lambda$ is established by
Eq.\ (\ref{eq:ldef}) [note that $M_{11}=M_{11}'$
in the rescaling Eq.\ (\ref{eq:resc})].
According to Eqs.\ (\ref{eq:dm}) and (\ref{eq:resc}), $u$ and $z$
fulfill the differential equations
\begin{equation}
\frac{du}{dt'}=z,\qquad\frac{dz}{dt'}=v'-z^2.
\label{eq:dgluz}
\end{equation}

Note that the evolution equation of $z$ decouples from $u$
and can be interpreted as a Langevin equation.
Hence the distribution $P(z;t')$ can be calculated from a Fokker-Planck
equation, which was
considered before in the context of wavefunction localization
\cite{Halperin,lifshitz},
\begin{subequations}
\label{eq:fpz}
\begin{eqnarray}
\partial_{t'} P(z;t') &=&  {\cal L}_z P(z;t'),\\
 {\cal L}_z \,\,\cdot & = & 
\partial_z(z^2+V_2'+\partial_z) \,\,\cdot.
\end{eqnarray}
\end{subequations}
For large $t'$ the distribution function $P(z;t')$ approaches
the stationary solution  \cite{Frisch,Halperin,lifshitz}
\begin{subequations}
\label{eq:pz}
\begin{eqnarray}
P_{\text{stat}}(z)&=&\tilde N \int_{-\infty}^z dy\,K(y,z),\\
K(y,z)&=&e^{(y^3-z^3)/3+ V_2'(y-z)},\\
\tilde N&=& \pi^{-2} [\ai^2(-V_2')+\bi^2(-V_2')]^{-1}.
\end{eqnarray}
\end{subequations}
Here $\ai$ and $\bi$ are Airy functions.
The normalization constant is directly related to the integrated
density of states
in the localization problem \cite{Frisch,Halperin,lifshitz}.
For $V_2'=0$, $\tilde N= 3^{5/6}2^{-1/3}\pi^{-1/2}/[\Gamma(1/6)]$.
Because $du/dt=z/t_c$ it is clear
\cite{lifshitz}
that the infinite-time Lyapunov exponent can
be obtained from $\lambda_\infty=\langle z\rangle/t_c$; this relation will be
demonstrated explicitely in Sec.\ \ref{sec:m1m2}.

The Fokker-Planck equation for the joint distribution
function $P(u,z;t')$ is given by
\begin{equation}
\partial_{t'} P = -z \partial_u P+{\cal L}_z  P.
\label{eq:fpuz}
\end{equation}
This Fokker-Planck
equation with $V_2'=0$ has been derived in Ref.\ \cite{al1} for the
autonomous chaotic scattering of a particle from a
dilute collection of scatterers (with more than one degree of freedom).

The joint distribution function $P(u,z;t')$ does not approach a stationary limit
because $u$ runs away to infinitely large values.
In order to analyze the behavior of the distribution function $P(u,z;t')$
for large times we convert the Fokker-Planck equation (\ref{eq:fpuz}) into
an eigenvalue problem which discriminates between the different time
scales involved in this evolution.
For this purpose, we 
introduce into Eq.\ (\ref{eq:fpuz}) the ansatz
\begin{equation}
P(u,z;t')=\int_{-i\infty}^{+i\infty}\frac{d\xi}{2\pi i}
\sum_{n=0}^\infty \exp(\mu_n t'-\xi u)f_n(\xi,z).
\label{eq:ansatz}
\end{equation}
(The integration contour along the imaginary axis
corresponds to an inverse  Laplace transformation.)
It follows that the functions $f_n$
fulfill the differential equation 
\begin{subequations}
\label{eq:eval}
\begin{equation}
\mu_n f_n(\xi,z)=(\xi z+{\cal L}_z)f_n(\xi,z),
\label{eq:eval1}
\end{equation}
in which $\mu_n$ and $\xi$ appear as parameters. However, in order to
obtain a meaningful probability distribution function (\ref{eq:ansatz})
we have to impose boundary conditions on $f_n(\xi,z)$ at $z\to \pm
\infty$. It is convenient to express these boundary conditions
by the requirement
\begin{equation}
{\cal P} \int_{-\infty}^\infty dz\,f_n(\xi,z) z < \infty.
\label{eq:eval2}
\end{equation}
\end{subequations}
Here ${\cal P}$ denotes the principal value with respect to the
integration boundaries at $\pm\infty$.
Condition (\ref{eq:eval2}) follows from the behavior $z\approx
(t'-t'_\infty)^{-1}$ of
the solution of the differential equation
(\ref{eq:dgluz}) close to times $t'\approx t'_\infty$
where $|z|\to\infty$ (and hence $v'$ can be ignored).
In practical terms, the
condition  (\ref{eq:eval2}) guarantees
that the drift of $u$ remains finite for all times.

Eqs.\ (\ref{eq:eval}) form an eigenvalue problem, 
since condition  (\ref{eq:eval2}) only can be fulfilled
for a discrete set of numbers $\mu_n$---note
that these eigenvalues depend on the parameter $\xi$.
In the limit of large $t'$ only the largest
eigenvalue $\mu_0(\xi)\equiv\mu(\xi)$ is relevant,
 because the other eigenvalues 
give rise to exponentially smaller contributions.
This eigenvalue vanishes as $\xi\to 0$,  {\em i.\,e.}, $\mu(0)=0$,
because the stationary distribution of $z$, Eq.\ (\ref{eq:pz}), 
must be recovered
for large times from Eq.\ (\ref{eq:ansatz}) by integrating out $u$.

The moments of $u$ are given by
\begin{eqnarray}
\langle u^n\rangle &=&\int_{-i\infty}^{+i\infty}\frac{d\xi}{2 \pi i}
\int_{-\infty}^\infty du\,
\exp(\mu t'-\xi u) f(\xi) u^n
\nonumber
\\ &=& \lim_{\xi\to 0}\partial_\xi^n\exp(\mu(\xi) t') f(\xi),
\label{eq:umom}
\end{eqnarray}
where the coefficients $f(\xi)=\int_{-\infty}^\infty dz f_0(\xi,z)$
are determined, in principle, by the initial condition for $P(u,z;t')$
at $t'=0$.
From Eq.\ (\ref{eq:umom}) we obtain the  moment-generating function 
\begin{equation}
\chi(\xi)=\langle \exp(\xi u)\rangle=\exp(\mu(\xi) t/t_c) f(\xi),
\label{eq:chi}
\end{equation}
where we re-introduced the original time variable $t=t_c t'$
by Eq.\ (\ref{eq:resc}).
The cumulant-generating function (\ref{eq:cumuldef}) hence takes the
form of Eq.\ (\ref{eq:res}), including the corrections of order $t^0$,
\begin{equation}
\eta(\xi)=\ln\chi(\xi)=\mu(\xi) t/t_c +\ln f(\xi).
\label{eq:cumul}
\end{equation}

The cumulants $\langle \langle \lambda^n \rangle \rangle$
of the finite-time Lyapunov exponent are obtained 
by expanding the generating function $\eta$ in powers of $\xi$, see Eq.\
(\ref{eq:cumuldef}).
In terms of the coefficients of the Taylor expansion
\begin{equation}
\mu=\sum_{n=1}^\infty \xi^n\mu^{(n)}
\label{eq:expanmu}
\end{equation}
[which
starts with the linear term in $\xi$
because $\mu(0)=0$], 
according to Eqs.\ (\ref{eq:cumuldef}) and (\ref{eq:cumul})
the $n$th cumulant of $\lambda$
is then given by
\begin{eqnarray}
\langle\langle \lambda^n\rangle \rangle= n! \mu^{(n)} t_c^{-1}  t^{1-n}
+{\cal O}(t^{-n}).
\label{eq:cumulexp}
\end{eqnarray}
This equation means that within the statistical model
the cumulants are universal quantities in the leading
order in $t$, in the sense that the initial conditions
$P(z,u;0)$ only enter the next-order
corrections.
The only system-specific parameters which enter the cumulants are the
time scale $t_c$ and the (rescaled) strength $V_2'$ of the static potential.
Note that ratios of cumulants are even independent of
the time scale $t_c$ (and hence of the parameters $D$ and $m$ of the statistical model).

The form (\ref{eq:tc}) of $t_c$ 
follows from Eq.\ (\ref{eq:cumulexp}) when $t_c$
is expressed in terms of the infinite-time Lyapunov exponent $\lambda_\infty$
with help of the definition 
\begin{equation}
\lambda_\infty=\lim_{t\to\infty}
\langle \lambda \rangle 
= \mu^{(1)}/t_c
.
\end{equation}
In terms of the bare quantities of the statistical model,
\begin{equation}
\lambda_\infty = \mu^{(1)} D^{1/3}m^{-2/3}.
\label{eq:lres}
\end{equation}

In the next two sections we obtain general expressions
for the expansion coefficients $\mu^{(n)}$ 
and calculate explicitely the proportionality factor
$\mu^{(1)}=d\mu/d\xi|_{\xi=0}$  in (\ref{eq:lres}), as well
as the first few coefficients $\mu^{(2)}$, $\mu^{(3)}$, $\ldots$,  which determine,
respectively, 
the variance and the leading non-Gaussian corrections (higher cumulants)
of the fluctuations of the finite-time
Lyapunov exponent around its limiting value $\lambda_\infty$.

\subsection{\label{sec:cumul}Recursion relations for the cumulants}

We now show how the cumulants can be calculated
from Eq.\ (\ref{eq:cumulexp})
by recursively solving a
hierarchy of equations for coefficients $\mu^{(n)}$
in the Taylor expansion of $\mu(\xi)$, Eq.\ (\ref{eq:expanmu}).

In analogy to  Eq.\ (\ref{eq:expanmu}) let us also
expand the function $f_0(\xi,z)$ in powers of $\xi$,
\begin{equation}
f_0(\xi,z)=\sum_{n=0}^\infty \xi^n f_0^{(n)}(z).
\label{eq:expanf}
\end{equation}
With Eqs.\ (\ref{eq:expanmu}) and (\ref{eq:expanf})
the eigenvalue problem (\ref{eq:eval}) can now be written order by order
in powers of $\xi^n$. For $n=0$ we recover the stationary variant
 (\ref{eq:fpz}) of the Fokker-Planck
equation (\ref{eq:fpuz}),
\begin{equation}
{\cal L}_z f_0^{(0)}(z)=0,
\label{eq:homogen}
\end{equation}
which is solved by the stationary solution $f_0^{(0)}(z)=P_{\text{stat}}(z)$,
Eq.\ (\ref{eq:pz}).
For $n>1$ the differential equations are of the form
\begin{equation}
{\cal L}_z f_0^{(n)}(z)
=-zf_0^{(n-1)}(z)
+\sum_{l=1}^n\mu^{(l)}f_0^{(n-l)}(z).
\label{eq:ordern}
\end{equation}

Let us assume that we have solved the hierarchy of equations up to order
$n-1$. 
In the next order $n$ both the unknown quantities $f_0^{(n)}$ as well as
$\mu^{(n)}$ appear.
The unknowns can be separated by integrating
the differential equation (\ref{eq:ordern})
over $z$ from $-\infty$ to $\infty$: The
integrated left-hand side vanishes because of condition (\ref{eq:eval2})
of the eigenvalue problem. The integrated right-hand side
can be rearranged to give $\mu^{(n)}$,
\begin{subequations}
\label{eq:recrels}
\begin{equation}
\mu^{(n)}=\int_{-\infty}^\infty dz \, [z f_0^{(n-1)}(z)
-\sum_{l=1}^{n-1}\mu^{(l)}f_0^{(n-l)}(z)],
\label{eq:murec}
\end{equation}
which only involves quantities up to order $n-1$.
Subsequently, $\mu^{(n)}$ can be inserted into Eq.\ (\ref{eq:ordern}).
The function
\begin{eqnarray}
f_0^{(n)}(z)&=&\int_{-\infty}^z dy \int_{-\infty}^y dx\,
K(y,z)
\nonumber\\
&& {}\times
[-x f_0^{(n-1)}(x)
+\sum_{l=1}^n\mu^{(l)}f_0^{(n-l)}(x)]
\quad
\label{eq:frec}
\end{eqnarray}
\end{subequations}
[with the kernel $K(y,z)$ defined in Eq.\ (\ref{eq:pz})]
is then obtained by solving the resulting inhomogeneous
differential equation with help of the partial solution
$f_0^{(0)}(z)$ of its homogeneous counterpart, Eq.\ (\ref{eq:homogen}).
This inhomogeneous
part of the  functions  $f_0^{(n)}(z)$ is fixed by the requirement that
$f_0^{(0)}(z)$ is normalized to 1. Adding the homogeneous
solution to $f_0^{(n)}(z)$ in  any order gives rise to additional
terms in all higher orders, but these combine in such a way 
that they drop out of the calculation of the coefficients $\mu^{(n)}$,
which hence are uniquely determined by Eq.\ (\ref{eq:murec}).

The recursion relations
(\ref{eq:recrels}) can be iterated to calculate successively all cumulants
of $\lambda$.

\begin{figure}
\begin{center}
\includegraphics[width=\columnwidth]{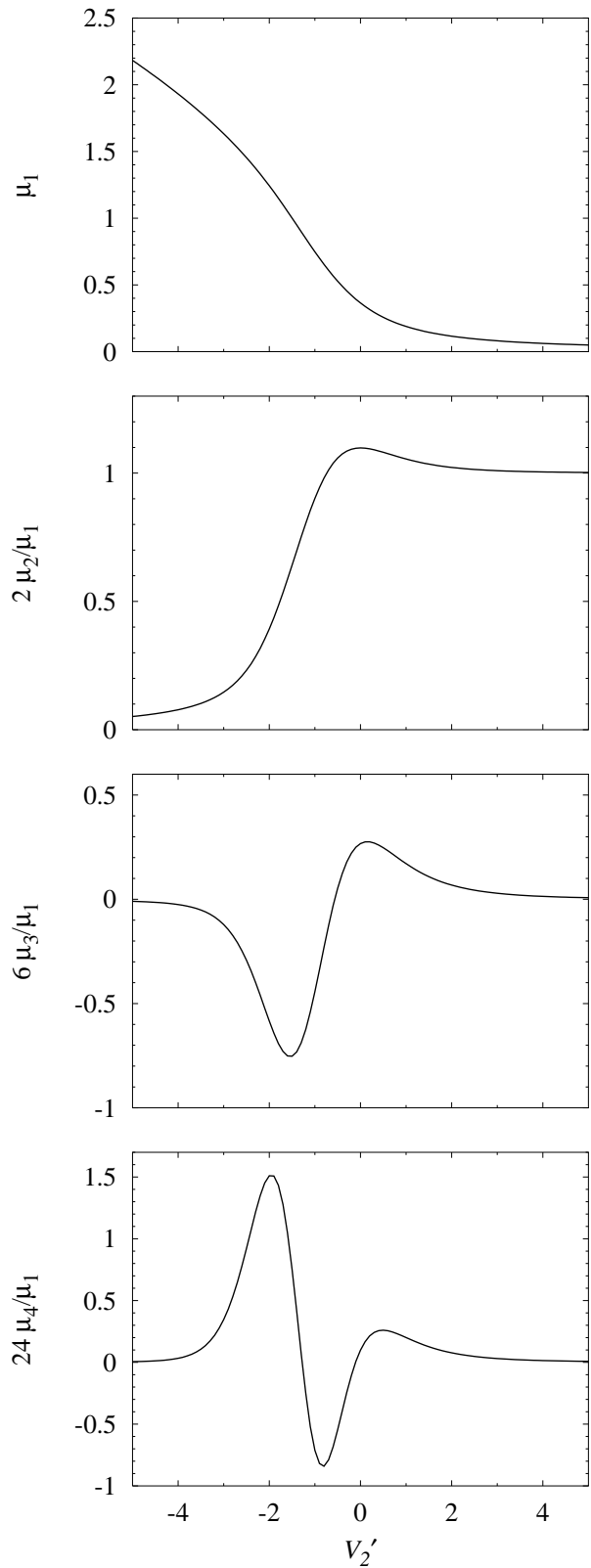}
\end{center}
\caption{Coefficient $\mu^{(1)}$ of the first cumulant, and the ratios $n!\mu^{(n)}/\mu^{(1)}$ for
the  coefficients of the second, third,
and fourth cumulant [cf.\ Eq.\ (\ref{eq:cumulexp})],
as a function of the strength $V_2'$ of the static background potential.}
\label{fig:fig1}
\end{figure}

\subsection{\label{sec:m1m2}Explicit expressions and numerical values}

According to Eq.\ (\ref{eq:cumulexp}), the
two numbers $\mu^{(1)}$ and $\mu^{(2)}$ determine
mean and variance
of the distribution function of $\lambda$, which then is approximated
by a Gaussian.
The coefficient $\mu^{(1)}$ has been obtained in Ref.\ \cite{lifshitz}
from the
Fokker-Planck equation (\ref{eq:fpz}) for arbitrary $V_2$.
For the special case $V_2=0$,
the two  coefficients $\mu^{(1)}$ and $\mu^{(2)}$ 
have been obtained in Ref.\ \cite{al1} from the
Fokker-Planck equation (\ref{eq:fpuz}).
However,
the deviations from the Gaussian distribution function are
not at all negligible for many chaotic systems,
which is most clearly displayed in their
generalized dimension and entropy spectra \cite{ott}.
As we have seen in the previous subsection \ref{sec:cumul},
our approach of reduction to the eigenvalue problem (\ref{eq:eval})
allows to analyze the non-Gaussian deviations
by the higher cumulants
of $\lambda$.
[In next section \ref{sec:pos},
we show that one can even obtain from our analysis
the positive moments of $M_{11}$, which are
determined by the far tail $\lambda \gg \lambda_\infty$
of $P(\lambda;t)$,
while the bulk of the distribution is essentially irrelevant
for these moments.]

Explicit expressions for the first few coefficients
$\mu^{(1)}$, $\mu^{(2)}$, $\mu^{(3)}$, and $\mu^{(4)}$ result
from Eq.\ (\ref{eq:murec}),
\begin{subequations}
\begin{eqnarray}
\mu^{(1)}&=&\int_{-\infty}^\infty dz \, z f_0^{(0)}(z),\\
\mu^{(2)}&=&\int_{-\infty}^\infty dz \, (z-\mu^{(1)}) f_0^{(1)}(z), \label{eq:m2int} \\
\mu^{(3)}&=&\int_{-\infty}^\infty dz \,
[(z-\mu^{(1)}) f_0^{(2)}(z)-\mu^{(2)}f_0^{(1)}(z)],\quad
\\
\mu^{(4)}&=&\int_{-\infty}^\infty dz \,
[(z-\mu^{(1)}) f_0^{(3)}(z)-\mu^{(2)}f_0^{(2)}(z)
\nonumber \\
&&{} \qquad
-\mu^{(3)}f_0^{(1)}(z)],
\end{eqnarray}
\end{subequations}
where  $ f_0^{(0)}(z)=P_{\text{stat}}(z)$ is given by 
the  stationary  distribution  function  of $z$, Eq.\
(\ref{eq:pz}), while the other functions
follow from Eq.\
(\ref{eq:frec}),
\begin{subequations}
\begin{eqnarray}
f_0^{(1)}(z)&=& 
\int_{z>y>x}\hspace*{-.8cm} dy\, dx\,
K(y,z)
(\mu^{(1)}-x)f_0^{(0)}(x)
,
\\
f_0^{(2)}(z)&=&\int_{z>y>x}\hspace*{-.8cm} dy\, dx\,K(y,z)
[(\mu^{(1)}-x)f_0^{(1)}(x)
\nonumber \\
&&\qquad{}-\mu^{(2)}f_0^{(0)}(x)]
,
\\
f_0^{(3)}(z)&=&\int_{z>y>x}\hspace*{-.8cm} dy\, dx\,K(y,z)
[(\mu^{(1)}-x)f_0^{(2)}(x)
\nonumber \\
&&\qquad{}-\mu^{(2)}f_0^{(1)}(x) -\mu^{(3)}f_0^{(0)}(x)]
.
\end{eqnarray}
\end{subequations}

The coefficient $\mu^{(1)}$ is then given by \cite{lifshitz}
\begin{eqnarray}
\mu^{(1)}
=\frac{1}{2}\frac{d}{dV_2'}\log\tilde N,
\end{eqnarray}
where $\tilde N$ is given in Eq.\ (\ref{eq:pz}),
while the cumulants for $n\geq 2 $
can be obtained quickly by numerical integration
of
$2n$-fold integrals. The effort of integration can be greatly reduced
down to the expense equivalent to a single integral, because the integrand
factorizes. An efficient recursive scheme is described in the Appendix.
In Fig.\ \ref{fig:fig1} we plot the coefficient
$\mu^{(1)}$ and the ratios $n! \mu^{(n)}/\mu^{(1)}$
for $n=2,3,4$ as a function of $V_2'$.
The non-Gaussian corrections are largest around $V_2'=0$, while
they become irrelevant for 
large negative or positive values of $V_2'$.

For strong confinement, $V_2'\gg 1$, the coefficients
$n!\mu^{(n)}/\mu^{(1)}\to\delta_{1n}+\delta_{2n}$, with $\delta_{mn}$
the Kronecker symbol, 
and the Gaussian approximation
\begin{equation}
\mu_{\text{Gaussian}}(\xi)=\mu^{(1)}\left(\xi + \frac{1}{2}\xi^2\right)
\label{eq:gaussian}
\end{equation}
becomes valid. [In the context of wave localization,
this corresponds to the well-known limit of a
large Fermi energy $E\sim V_2'$ (cf.\ Sec.\ \ref{sec:loc}).]

\begin{table}
\begin{ruledtabular}
\begin{tabular}{|c|c|c|c|c|}
\hline
$n $&1&2&3&4\\
\hline
$n!\mu^{(n)}$
& $0.365$
& $0.401$
& $0.0975$
& $0.0361$
\\
\hline
\hline
$n $&5 &6&7&8\\
\hline
$n!\mu^{(n)}$
& $-0.266$
& $-0.628$
& $-0.554$
& $3.71$
\\
\hline
\end{tabular}
\end{ruledtabular}
\caption{First eight coefficients $n!\mu^{(n)}$
of the cumulants of finite-time Lyapunov exponents  [cf.\ Eq.\ (\ref{eq:cumulexp})], 
in absence of the static background potential ($V_2'=0$).}
\label{tab:tabv0}
\end{table}

Analytical results can be found in the case $V_2'=0$ for the first two
coefficients,
\begin{subequations}
\begin{eqnarray}
\mu^{(1)}
&=&
\frac{(3/2)^{1/3}\sqrt{\pi}}{\Gamma(1/6)}
,
\\
\mu^{(2)}&=&
\frac{5\pi^2}{18} \tilde N -\frac{\pi}{2\sqrt{3}}\tilde N \,
{}^{}_3F_2\left( 1,1, \frac{7}{6}; \frac{3}{2},\frac{3}{2};
\frac{3}{4}\right)
,\quad
\label{eq:m2}
\end{eqnarray}
\end{subequations}
where $\tilde N(V_2'=0)=
3^{5/6}2^{-1/3}\pi^{-1/2}/[\Gamma(1/6)]$, while $_3F_2$ is
a generalized hypergeometric function. 
Incidentally,
the numerical value given for $\mu^{(2)}$ in Ref.\ \cite{al1} is
wrong, but the analytic expression given in that paper
is equivalent to Eqs.\ (\ref{eq:m2int}) and (\ref{eq:m2}).
In Tab.\ \ref{tab:tabv0} we tabulate
the numerical
values of the first eight coefficients $n!\mu^{(n)}$ for $V_2'=0$.

\section{\label{sec:pos}Positive moments}

\subsection{\label{sec:genmom}Formally exact expressions}

In view of 
Eqs.\ (\ref{eq:u}) and  (\ref{eq:chi}) we find that the exponential
growth rates
of the positive moments of $M_{11}$ are given by the eigenvalue $\mu(\xi)$
of Eq.\ (\ref{eq:eval}) at integer values of $\xi$:
\begin{equation}
\frac{d\ln\langle M_{11}^\xi \rangle}{d t}=\frac{\mu(\xi)}{t_c}
.
\label{eq:growthrates}
\end{equation}
As we will now show, for integer values of $\xi$ the eigenvalue problem
(\ref{eq:eval}) can be reduced to a matrix eigenvalue problem of finite
dimension. For the first few moments the leading eigenvalue can be
calculated explicitely, while for larger values it is formally given by
the largest root of the corresponding characteristic polynomial.

\begin{table}
\begin{ruledtabular}
\begin{tabular}{|c|c|c|c|c|}
\hline
$\xi $&1&2&3&4\\
\hline
$\mu$
& $0$
& $2^{2/3}$
& $ 24^{1/3}$
& $84^{1/3}$
\\
\hline
\hline
$\xi $&\multicolumn{2}{c|}{5} &6&7\\
\hline
$\mu$
& \multicolumn{2}{c|}{$2(14+3\sqrt{19})^{1/3}$}
& $(252+24\sqrt{79})^{1/3}$
& $ 2 (63+15\sqrt{10})^{1/3}$
\\
\hline
\end{tabular}
\end{ruledtabular}
\caption{Exponential growth rates $\mu(\xi)$ of the
first few moments $\langle M_{11}^\xi\rangle$ [cf.\ Eq.\ (\ref{eq:growthrates})],
in absence of the static background potential ($V_2'=0$).}
\label{tab:1}
\end{table}

In order to obtain a solution
of the differential equation (\ref{eq:eval1}),
we write
\begin{equation}
f_0(\xi,z)=\int_{-\infty}^z dy\,K(y,z)\frac{g(z)}{g(y)^{2}}
\end{equation}
[with the kernel $K(y,z)$ defined in Eq.\ (\ref{eq:pz})],
 and obtain for $g$ the differential equation
\begin{equation}
(\mu-\xi z)g=-(z^2+V_2') \partial_z g+ \partial_z^2 g
\label{eq:g}
\end{equation}
(a triconfluent Heun's equation with singularity at $1/z=0$).
We introduce into this equation the polynomial ansatz
\begin{equation}
g(z)=\sum_{n=0}^\xi c_n z^n.
\label{eq:g1}
\end{equation}
Power matching results in
the following recursion relation
\begin{subequations}
\label{eq:recrel}
\begin{equation}
(\xi-n)c_n=\mu c_{n+1}+(n+2)[V_2'c_{n+2} -(n+3)c_{n+3}]
\end{equation}
for the  coefficients $c_n$, with initial conditions
\begin{equation}
c_\xi=1,\qquad c_{\xi-1}=\mu,\qquad c_{\xi-2}=\mu^2/2.
\end{equation}
\end{subequations}
For integer $\xi$ this
recursion relation terminates.
We 
obtain functions $c_0(\mu)$,  $c_1(\mu)$,  and $c_2(\mu)$ and an
additional condition from the term in Eq.\ (\ref{eq:g})
which is constant in $z$,
\begin{equation}
p_\xi(\mu)=\mu c_0+V_2'c_1-2c_2=0,
\end{equation}
where $p_\xi(\mu)$ is a polynomial of degree $\xi+1$.

The polynomial $p_\xi(\mu)$ can also be interpreted as the characteristic
polynomial of the $(\xi+1)\times (\xi+1)$-dimensional matrix 
\begin{equation}
\left(
\begin{array}{cccccccc}
 0& -V_2' & 1\cdot 2 & 0 & 0 & \cdots   & \cdots & \cdots\\
\xi & 0 & -2 V_2' & 2\cdot 3 & 0 & \cdots  & \cdots & \cdots\\
0 & \xi-1 & 0 & -3 V_2' & 3\cdot 4 & \cdots   & \cdots & \cdots\\
0 & 0 & \xi-2 & 0 & -4 V_2' &  \cdots   & \cdots & \cdots\\
0 & 0 & 0 & \xi-3 & 0 &  \ddots   & \cdots & \cdots\\
\vdots &\vdots &\vdots &\vdots &\ddots & 0 & (1-\xi)V_2' & (\xi-1)\xi \\
\vdots &\vdots & \vdots &\vdots &\vdots & 2 & 0 &  -\xi V_2'  \\
\vdots &\vdots & \vdots &\vdots & \vdots  & 0 & 1 &  0 
\end{array}
\right)
,
\label{eq:mat}
\end{equation}
which is the matrix representation of the eigenvalue problem
(\ref{eq:eval}) in  the space of the monomial expansion of $g(z)$.

The exponential growth
rate $\mu(\xi)$ of the $\xi$th moment is given by 
the largest root of $p_\xi(\mu)$, or equivalently by
the largest eigenvalue of the matrix 
(\ref{eq:mat}). 
In subsection \ref{sec:direct} we will see for the examples $\xi=1,2$
that the other roots show up in the transient behavior of the moments.

\begin{figure}
\begin{center}
\includegraphics[width=0.4\textwidth]{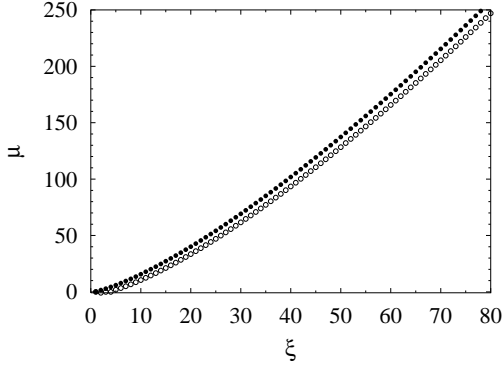}
\end{center}
\caption{
Growth rates $\mu(\xi)$ of the  moments $\langle M_{11}^m\rangle$
[cf.\ Eq.\ (\ref{eq:growthrates})] in absence of the static background potential
($V_2'=0$),
obtained as the largest eigenvalue of the matrix  (\ref{eq:mat}) 
(full circles). Also shown is the real part of
the subleading eigenvalue of this matrix (open
circles).
}
\label{fig:fig2}
\end{figure}

First we present results in absence of the static background potential,
$V_2'=0$. 
The values
for the first few moments are given in Tab.\ \ref{tab:1}. 
Figure \ref{fig:fig2} shows the growth rates and the 
real part of the subleading eigenvalue
for values of $\xi$ up to $80$. A log-normal statistics of $M_{11}$
(corresponding to a Gaussian statistics of the finite-time Lyapunov
exponents) would result in the quadratic dependence Eq.\
(\ref{eq:gaussian}) of $\mu(\xi)$ on
$\xi$, 
while the plot shows a weaker (approximately linear) dependence for large $\xi$.
This results from the influence of the terms $\mu^{(n)}\xi^n$ for $n\geq 3$ in the complete
Taylor expansion of $\mu$, Eq.\ (\ref{eq:expanmu}).
Further
note that the subleading eigenvalue stays at a finite distance to the leading
eigenvalue (indeed, their distance increases with increasing $\xi$),
as we have assumed before in restricting the attention
to the leading eigenvalue $\mu_0$ of the
eigenvalue problem (\ref{eq:eval}).

\begin{figure}
\begin{center}
\includegraphics[width=\columnwidth]{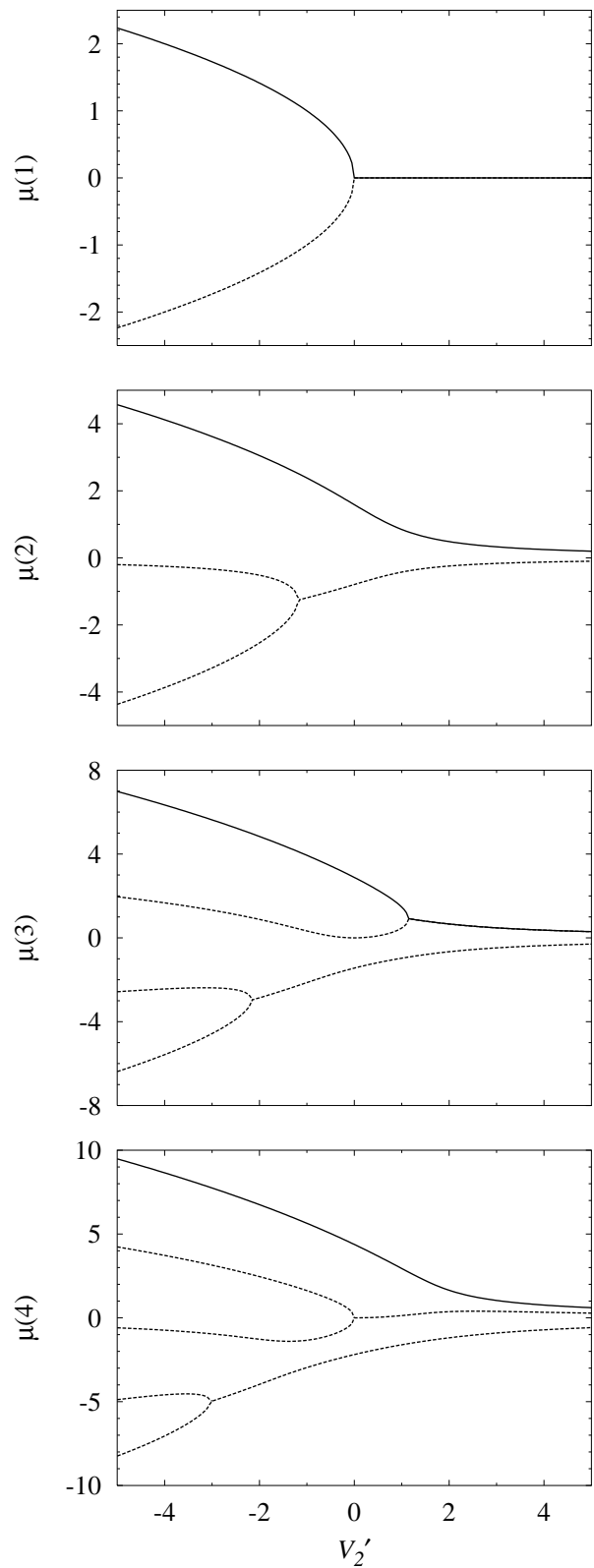}
\end{center}
\caption{
Growth rates $\mu(\xi)$ the  moments $\langle M_{11}^m\rangle$ [cf.\ Eq.\ (\ref{eq:growthrates})], for $m=1,2,3,4$,
as a function of the strength $V_2'$ of the static background potential.
Also shown (dashed lines) is the real part of
the subleading growth rates [subleading eigenvalues of  matrix (\ref{eq:mat})].
}
\label{fig:fig3}
\end{figure}

For finite $V_2'$, the growth rate of the first moment 
\begin{equation}
\re \mu(1)=|\im\sqrt{V_2'}|
\end{equation}
vanishes in the case of confinement, $V_2'>0$.
This will be confirmed by the direct computation in Sec.\ \ref{sec:direct}.
The growth rate of the second moment is given by
\begin{eqnarray}
\mu(2)&=&2^{1/3}\left(
1+\sqrt{1+16 {V_2'}^3/27}\right)^{1/3}
\nonumber\\
&&{}+2^{1/3}\left(1-\sqrt{1+16 {V_2'}^3/27}\right)^{1/3}
\end{eqnarray}
[with the roots taken such that $\mu(2)$ is real].
We plotted the real parts of the leading and subleading growth rates 
[eigenvalues of 
matrix (\ref{eq:mat})] for the first four moments in Fig.\ \ref{fig:fig3}. 

\subsection{\label{sec:direct}Direct computation of the first and second moment}

In order to illustrate the results for the growth rates
of the moments $\langle M_{11}^\xi\rangle$
we compare the results for $\xi=1$ and $\xi=2$ to the exact
results for all times (including the transient behavior).
A formal solution of the differential
equation (\ref{eq:m11eq}) in terms of a series in the disorder potential
is obtained by integrating Eq.\ (\ref{eq:m11eq}) twice, under
observation of the initial conditions $M_{11}=1$, $d M_{11}/dt=0$
for $t=0$, and iterating  the resulting integral relation
\begin{eqnarray}
M_{11}(t)&=&1+\int_0^{t}dt_1\int_0^{t_1} ds_1\frac{v(s_1)}{m}M_{11}(s_1)
\nonumber
\\
&=&1+\int_0^{t}dt_1(t-t_1)\frac{v(t_1)}{m}M_{11}(t_1).
\end{eqnarray}
The formal solution is of the form
\begin{equation}
M_{11}(t_0)=1+\sum_{n=1}^\infty
\prod_{k=1}^n\int_0^{t_{k-1}}
dt_k\, (t_{k-1}-t_k) \frac{v(t_k)}{m},
\label{eq:m11}
\end{equation}
where we introduced $t_0=t$ for notational convenience.

For the first moment we can average Eq.\ (\ref{eq:m11}) directly.
Because of the factors $ (t_{k-1}-t_k)$ and the time ordering, the 
random function $v$ never appears instantaneously
in second or higher order in any of the
integrals. Hence we can replace $v$ by its average, given in
Eq.\ (\ref{eq:v}).
Consequently, the first moment is given by
\begin{equation}
\langle M_{11}\rangle = \cos[(t/t_c)\sqrt{V_2'}] =
\frac{1}{2}e^{(t/t_c)\sqrt{-V_2'}}+\frac{1}{2}e^{-(t/t_c)\sqrt{-V_2'}}.
\end{equation}
For $V_2'=0$ the first moment is
constant and given by its
initial value,
$\langle M_{11}\rangle=1$.
This means that negative deviations $M_{11}\ll 0$,
corresponding to inverse hyperbolic motion, cancel
precisely the
positive deviations $M_{11}\gg 0$ of hyperbolic motion.
For negative $V_2'$ the first moment grows, while for positive
$V_2'$ it oscillates and stays of order unity.
In the decomposition of the cosine into the two exponentials,
we identify in the exponents
the two  roots $\pm \sqrt{-V_2'}$ of the characteristic polynomial
$p_{\xi=1}(\mu)=\mu^2+V_2'$ of the matrix
(\ref{eq:mat}) with $\xi=1$. For negative $V_2'$, the subleading exponent
hence governs the transient behavior of the first moment.

For the second moment let us restrict for simplicity to the case $V_2'=0$.
We group the
functions $v$ in the two factors of $M_{11}$ in pairs and then invoke the
delta-correlations of  Eq.\ (\ref{eq:v}). Performing the time-ordered
integrals we obtain
\begin{eqnarray}
&&\langle M_{11}^2\rangle =
1+\sum_{n=1}^\infty (2/t_c^3)^n\prod_{k=1}^n\int_0^{t_{k-1}}
dt_k\, (t_{k-1}-t_k)^2 
\nonumber \\
&&= \frac{1}{3} [e^{\mu(2)t/t_c}+
e^{-(-1)^{1/3}\mu(2)t/t_c}+
e^{-(-1)^{-1/3}\mu(2)t/t_c}].
\nonumber \\
\label{eq:m11trans}
\end{eqnarray}
The asymptotic 
growth rate of the second moment is given by the leading root $\mu(2)=2^{2/3}$
of the characteristic polynomial $p_2(\mu)=\frac{1}{2}\mu^3-2$,
which is in accordance to Tab.\ \ref{tab:1}.
The second and third exponent are the other two roots of this polynomial.

\subsection{\label{sec:equiv}Equivalence of matrix elements}

So far we mainly studied the statistics of the upper diagonal element 
$M_{11}$ of the stability matrix $M$.
At this point now we can discuss how the results for the
cumulant-generating function and the positive moments can be transferred
to the other elements of $M$.

The differential equation (\ref{eq:m2112})
for $M_{22}$ can be integrated similarly as
the one for $M_{11}$, 
from which we obtain analogously to Eq.\ (\ref{eq:m11}) the formal solution
\begin{equation}
M_{22}(t_0)=1+\sum_{n=1}^\infty
\prod_{k=1}^n\int_0^{t_{k-1}}
dt_k\, (t_{k}-t_{k+1}) \frac{v(t_k)}{m}.
\label{eq:m22}
\end{equation}
Here we defined in each term of order $n$ that $t_{n+1}=0$.
It follows by direct computation that the
first two moments of $M_{22}$ are identical to those of $M_{11}$,
\begin{equation}
\langle M_{22} \rangle=\langle M_{11}\rangle,\quad
\langle M_{22}^2 \rangle=\langle M_{11}^2\rangle.
\end{equation}
These explicit results 
already suggest that the
statistics of the two diagonal matrix elements is the same. Indeed,
the transformation $t_k=t-\tilde{t}_{n+1-k}$,
$v(t-\tilde{t})=\tilde{v}(\tilde{t})$ brings Eq.\ (\ref{eq:m22}) into
the form of Eq.\ (\ref{eq:m11})
and leaves the properties of the Gaussian noise (\ref{eq:v}) invariant.
Hence even the transient behavior of the diagonal elements is
completely identical, for arbitrary values of $V_2'$.

The results for the cumulant-generating function $\eta(\xi)$
(hence also the growth rates of the moments, but not the transient
behavior)
can also be transferred to the offdiagonal matrix elements
of $M$: The element
$M_{12}$ fulfills the same differential equation as $M_{11}$,
see Eq.\ (\ref{eq:m11eq}),
while $M_{21}$ fulfills the same differential equation as $M_{22}$.
The initial conditions of the offdiagonal matrix elements
differ from those of the diagonal elements.
However, according to Eq.\ (\ref{eq:cumul})
this only affects the function $f(\xi)$ in the subleading
corrections of the cumulant-generating function
[which, for the example of the second moment, results in
factors in front of the exponential functions
which are different than
in Eq.\ (\ref{eq:m11trans})]. 

Let us add that from Eqs.\ (\ref{eq:m11}) and (\ref{eq:m22})
we find  for $V_2'=0$ the cross-correlator
\begin{equation}
\langle M_{11}M_{22}\rangle=\frac{1}{2}+\frac{1}{2}\langle M_{11}^2\rangle.
\end{equation}
As a consequence, for $V_2'=0$
the trace $\tr M=M_{11}+M_{22}$ of the stability matrix has the following
first two moments
\begin{subequations}
\begin{eqnarray}
\langle\tr M \rangle& =&
2,
\\
\langle( \tr M)^2 \rangle
&=& 1+e^{\mu(2)t/t_c}+2 \,\re
e^{-(-1)^{1/3}\mu(2)t/t_c}
.
\nonumber\\
\end{eqnarray}
\end{subequations}

\section{\label{sec:dis}Discussion}

In this work we presented a uniform approach to the asymptotic statistics 
of finite-time  Lyapunov exponents, for the model
(described in Sec.\ \ref{sec:form})
of a particle moving in a random time-dependent
potential. The cumulant-generating function $\eta(\xi)$ 
was found to be directly proportional to the eigenvalue $\mu(\xi)$ of a
parameterized differential equation, defined by Eqs.\ (\ref{eq:eval}).
This facilitated an effective analysis of the statistics,
including the non-Gaussian deviations of the distribution function.
These deviations are especially important
for the positive moments of the elements
of the stability matrix, since their growth
rate {\em  cannot} be predicted by the Gaussian approximation
Eq.\ (\ref{eq:gaussian}).

We limited our attention to the case 
of time-dependent Hamiltonian systems with 
a single degree of freedom and a
Hamiltonian (\ref{eq:ham})
which is of the special type of kinetic energy
plus potential energy, with
time-dependence only in the potential energy.
This case is of particular interest because of its direct applicability to specific
dynamical systems as in the random wave-propagation problem of
Refs.\ \cite{wolfson,brown},
and  because of its applicability to one-dimensional wave localization.
For the Hamiltonian (\ref{eq:ham}) the matrix $K$ in the differential
equation (\ref{eq:dm}) is purely off-diagonal,
with fluctuations only in the lower-left
element.
For Hamiltonians which do not separate into kinetic and potential
energy, the differential
equation (\ref{eq:ham}) for $M$ involves the matrix $K$
in the more general form
\begin{equation}
K=
\left(\begin{array}{cc}
K_{11} &K_{12}\\
K_{21} &K_{22}
\end{array}\right)
=
\left(\begin{array}{cc}
\frac{\partial^2 H}{\partial x\partial p} & \frac{\partial^2 H}{\partial p^2} \\
-\frac{\partial^2 H}{\partial x^2} & - \frac{\partial^2 H}{\partial x\partial p}
\end{array}\right)
.
\label{eq:kgen}
\end{equation}
A generalized statistical model now arises
by introducing noise into all of the matrix elements of $K$.
(One may also allow for correlations between the
different matrix elements or for finite correlation times by introducing auxiliary variables 
for the noise in the standard way.)

Let us point out two particular cases for
which a statistical description promises
to result in  direct applications to physical situations
of interest.
One case is more
relevant to wave-function localization while the other is more relevant for chaotic
dynamics.

a) The diagonal elements $K_{11}=-K_{22}=0$  still vanish identically, but both 
off-diagonal elements $K_{12}$ and
$K_{21}$ fluctuate with a vanishing mean. This situation appears to be
related
to the band-center case of
one-dimensional localization in the  Anderson model \cite{kappus,goldhirsch}
(where space is
discretized on the lattice), since at the band-center the
effective mass of the
particle diverges (and hence the mean of $K_{12}$ vanishes).

b) Chaotic dynamics with an isotropic phase space may be modeled by
independent fluctuations of all four matrix elements $K_{ij}$ with identical
amplitude and vanishing mean.
Hamiltonian dynamics gives rise to the further constraint $K_{11}=-K_{22}$.
Isotropic dynamics arises in typical chaotic maps (some maps,
like the Baker map or the cat map,
however, are not isotropic---the directions of
stable and unstable manifolds are known by construction).
Good candidates are the Poincar{\'e} surface of section  of
autonomous systems with two degrees of freedom, 
in which the motion in four-dimensional
phase space is restricted to three-dimensional manifolds 
of constant energy and the coordinate 
along the flow field is taken as a time.

It would be interesting to compare the outcome of an analysis of 
model b)
with the  findings in the literature  \cite{Prasad,Diakonos}
which indicate a certain degree of robustness (if not universality)
of the distribution of finite-time  Lyapunov exponents for
typical chaotic systems.

\begin{acknowledgments}
We gratefully acknowledge useful discussions with Philippe Jacquod and
Holger Kantz, and especially with Steven Tomsovic who motivated us to
study this problem.
\end{acknowledgments}

\appendix 
\section*{Appendix: Integrals for the higher cumulants}
The cumulants of order $n$ result from the recursion relations Eq.\ (\ref{eq:recrels})
in the form of $2n$-fold integrals. Usually, the numerical
evaluation of such integrals for large $n$ is very time-consuming, since the number of points on a
grid covering the integration domain 
with lattice constant $(1/N)$, $N\gg 1$,
grows rapidly with $n$ as $N^{2n}$. However, presently the integrand factorizes and
the expense of the integration can be reduced from exponential to algebraic $n$-dependence  $\sim n N$. 
The principle can be demonstrated for the example of the two-fold integral
\begin{equation}
I^{(1)}= \int_{-z_0}^{z_1} dz\, I^{(2)}(z),\qquad I^{(2)}(z)=g(z)\int_{-z_0}^z dy\, I^{(3)}(y),
\end{equation}
where
$g$ is an arbitrary function and $I^{(3)}$ may itself be a multi-dimensional integral.

We introduce an index $m$ which denotes that the argument of a function is taken at the
$m$th lattice point on the appropriate axis of the grid.
The initial values of $I^{(n)}_m$ at $m=0$ (the lower integration boundary) are zero.
We now can write recursively, by incrementally increasing the integration variables,
\begin{subequations}
\label{eq:reci1i2}
\begin{eqnarray}
\label{eq:reci2}
I^{(2)}_{m+1}&=&\frac{g_{m+1}}{g_m} I^{(2)}_m+\frac{1}{N}g_mI^{(3)}_m,\\
I^{(1)}_{m+1}&=&I^{(1)}_m+\frac{1}{N} I^{(2)}_{m+1}.
\label{eq:reci1}
\end{eqnarray}
\end{subequations}
Moreover,
when $I^{(3)}$ itself is a multi-dimensional integral of type $I^{(1)}$,
its current value can be obtained recursively in the same way as the value of $I^{(1)}$.
Since each additional integral will give rise to
only one additional equation [similar either to Eq.\ (\ref{eq:reci2})
or to Eq.\ (\ref{eq:reci1})], the number of operations grows linearly with $n$, as advertised above.
[The recursion relations (\ref{eq:reci1i2})
have the additional advantage for the present problem that
they  avoid over- and underflow in the evaluation of the kernel $K(y,z)=\exp(y^3/3+V_2' y-z^3/3-V_2' z)$.]

\end{document}